\documentclass{JAC2001}

\usepackage{graphicx}
\newcommand{\ud}{\mathrm{d}}
\begin{document}

\title{COHERENT STRUCTURES AND PATTERN FORMATION IN VLA\-SOV\--MAX\-WELL\--PO\-ISSON SYSTEMS}
\author{Antonina  N. Fedorova, Michael  G. Zeitlin \\
IPME, RAS, V.O. Bolshoj pr., 61, 199178, St.~Petersburg, Russia
\thanks{e-mail: zeitlin@math.ipme.ru}\thanks{ http://www.ipme.ru/zeitlin.html;
http://www.ipme.nw.ru/zeitlin.html}   
}
\maketitle

\begin{abstract}
We present the applications of methods from nonlinear local harmonic analysis      
for calculations in nonlinear collective dynamics           
described by different forms of Vlasov-Maxwell-Poisson equations. Our           
approach is based on methods provided                                          
the possibility to work with well-localized in phase space bases, which gives     
the most sparse representation for the general type of operators and good      
convergence properties. The consideration is based on                              
a number of anzatzes, which reduce initial problems to a number of             
dynamical systems and on variational-wavelet approach to                       
po\-ly\-no\-mi\-al\-/ra\-ti\-o\-nal approximations for nonlinear dynamics. 
This approach  allows us to                                                                   
construct the solutions via nonlinear high-localized  eigenmodes               
expansions in the base of compactly supported wavelet bases and control        
contribution from each scale of underlying multiscales. Numerical              
modelling demonstrates formation of coherent structures and stable             
patterns.                                                                      
\end{abstract}

\section{INTRODUCTION}
In this paper we consider the applications of nu\-me\-ri\-cal\--analytical 
technique based on the methods of local nonlinear harmonic
analysis or wavelet analysis to nonlinear beam/accelerator physics
problems which can be characterized by collective type behaviour and described by 
some forms of Vlasov-Maxwell-Poisson equations [1].
Such approach may be useful in all models in which  it is 
possible and reasonable to reduce all complicated problems related with 
statistical distributions to the problems described 
by systems of nonlinear ordinary/partial differential 
equations with or without some (functional)constraints.
Wavelet analysis is a set of mathematical
methods, which gives the possibility to work with well-localized bases in
functional spaces and gives the maximum sparse forms for the general 
type of operators (differential, integral, pseudodifferential) in such bases. 
Our approach is based on the 
variational-wavelet approach from [2]-[13],
which allows us to consider polynomial and rational type of 
nonlinearities.
The solution has the multiscale/multiresolution decomposition via 
nonlinear high-localized eigenmodes, 
which corresponds to the full multiresolution expansion in all underlying time/space 
scales. 
The same is correct for the contribution to power spectral density
(energy spectrum): we can take into account contributions from each
level/scale of resolution.
In all these models numerical modelling demonstrates the appearance of coherent high-localized structures
and stable patterns formation.
Starting  from Vlasov-Maxwell-Poisson equations in part 2,
we consider the approach based on
variational-wavelet formulation in part 3. 
We give the explicit representation for all dynamical variables in the base of
compactly supported wavelets or nonlinear eigenmodes.  Our solutions
are parametrized
by solutions of a number of reduced algebraical problems one from which
is nonlinear with the same degree of nonlinearity as initial differential problem and the others  are
the linear problems which correspond to the particular
method of calculations inside concrete wavelet scheme. 
In part 4 we consider numerical modelling based on our analytical approach.

\section{COLLECTIVE MODELS VIA
VLASOV-MAXWELL-POISSON EQUATIONS}

Analysis based on the non-linear Vla\-sov\--Max\-well\--Poi\-s\-son equations leads to more
clear understanding  collective effects and nonlinear beam dynamics
of high intensity beam propagation in pe\-ri\-o\-dic\--fo\-cu\-sing and uni\-form\--fo\-cu\-sing
transport systems.
We consider the following form of equations (ref. [1] for setup and designation): 

{\setlength\arraycolsep{0pt}
\begin{eqnarray}
&&\Big\{\frac{\partial}{\partial s}+p_x\frac{\partial}{\partial x}+
             p_y\frac{\partial}{\partial y}-
\Big[k_x(s)x+\frac{\partial\psi}{\partial x}\Big]\frac{\partial}{\partial p_x}-\nonumber\\
&& \Big[k_y(s)y+\frac{\partial\psi}{\partial y}\Big]\frac{\partial}{\partial p_y}
  \Big\} f_b(x,y,p_x,p_y,s)=0, \\
&&\Big(\frac{\partial^2}{\partial x^2}+\frac{\partial^2}{\partial y^2}\Big)\psi=
-\frac{2\pi K_b}{N_b}\int \ud p_x \ud p_y f_b,\\
&&\int\ud x\ud y\ud p_x\ud p_y f_b=N_b
\end{eqnarray}}
The corresponding Hamiltonian for transverse sing\-le\--par\-ticle motion is given by 
{\setlength\arraycolsep{0pt}
\begin{eqnarray}
&& H(x,y,p_x,p_y,s)=\frac{1}{2}(p_x^2+p_y^2) 
                   +\frac{1}{2}[k_x(s)x^2 \\
 &&+k_y(s)y^2]+
    H_1(x,y,p_x,p_y,s)+\psi(x,y,s), \nonumber
\end{eqnarray}}
where $H_1$ is nonlinear (polynomial/rational) part of the full Hamiltonian.
In case of Vlasov-Maxwell-Poisson system we may transform (1) into invariant form
\begin{eqnarray}
\frac{\partial f_b}{\partial s}+[f_b,H]=0.
\end{eqnarray}

\section{MULTISCALE REPRESENTATIONS}

We obtain our multiscale/multiresolution representations (formulae (11) below) for solutions of equations
(1)-(5) via variational-wavelet approach for the following formal systems of equations (with
corresponding obvious constraints on distribution function),which are the general form of these equations. 
Let L be an arbitrary (non) \-li\-ne\-ar dif\-fe\-ren\-tial\-/in\-teg\-ral operator with matrix dimension $d$, 
which acts on some set of functions
$\Psi\equiv\Psi(s,x)=\Big(\Psi^1(s,x),\dots,\Psi^d(s,x)\Big)$, $ s,x \in\Omega\subset{\bf R}^{n+1}$
from $L^2(\Omega)$:
\begin{equation}
L\Psi\equiv L(R(s,x),s,x)\Psi(s,x)=0,
\end{equation}
($x$ are the generalized space coordinates or phase space coordinates, $s$ is ``time'' coordinate).
After some anzatzes [13],[14] the main reduced problem may be formulated as the system of ordinary differential            
equations                                                              
{\setlength\arraycolsep{0pt}                                                                
\begin{eqnarray}\label{eq:pol0}                                
& & Q_i(f)\frac{\ud f_i}{\ud s}=P_i(f,s),\quad f=(f_1,..., f_n),\\
& &i=1,\dots,n, \quad                                                                        
 \max_i  deg \ P_i=p, \quad \max_i deg \  Q_i=q \nonumber                  
\end{eqnarray}} 
\noindent or a set of such systems corresponding to each independent coordinate
in phase space. 
They have the fixed initial (or boundary) conditions $f_i(0)$, where $P_i, Q_i$ are not more    
than polynomial functions of dynamical variables $f_j$                                 
and  have arbitrary dependence on time. 
As result
we have the following reduced algebraical system
of equations on the set of unknown coefficients $\lambda_i^k$ of
localized eigenmode expansion (formula (9) below):
\begin{eqnarray}\label{eq:pol2}
L(Q_{ij},\lambda,\alpha_I)=M(P_{ij},\lambda,\beta_J),
\end{eqnarray}
where operators L and M are algebraization of RHS and LHS of initial problem
(\ref{eq:pol0}) and $\lambda$ are unknowns of reduced system
of algebraical equations (RSAE) (\ref{eq:pol2}).
After solution of RSAE (\ref{eq:pol2}) we determine
the coefficients of wavelet expansion and therefore
obtain the solution of our initial problem.
It should be noted if we consider only truncated expansion with N terms
then we have from (\ref{eq:pol2}) the system of $N\times n$ algebraical equations
with degree $\ell=max\{p,q\}$
and the degree of this algebraical system coincides
with degree of initial differential system.
So, we have the solution of the initial nonlinear
(rational) problem  in the form
\begin{eqnarray}\label{eq:pol3}
f_i(s)=f_i(0)+\sum_{k=1}^N\lambda_i^k f_k(s),
\end{eqnarray}
where coefficients $\lambda_i^k$ are the roots of the corresponding
reduced algebraical (polynomial) problem RSAE (\ref{eq:pol2}).
Consequently, we have a parametrization of solution of initial problem
by the solution of reduced algebraical problem (\ref{eq:pol2}).
The obtained solutions are given
in the form (\ref{eq:pol3}),
where
$f_k(t)$ are basis functions obtained via multiresolution expansions and represented by
some compactly supported wavelets.
Because affine
group of translation and dilations is inside the approach, this
method resembles the action of a microscope. We have contribution to
final result from each scale of resolution from the whole
infinite scale of spaces. More exactly, the closed subspace
$V_j (j\in {\bf Z})$ corresponds to  level j of resolution, or to scale j.
We consider  a multiresolution analysis of $L^2 ({\bf R}^n)$
(of course, we may consider any different functional space)
which is a sequence of increasing closed subspaces $V_j$:
$$
...V_{-2}\subset V_{-1}\subset V_0\subset V_{1}\subset V_{2}\subset ...
$$
satisfying the following properties:
let $W_j$ be the orthonormal complement of $V_j$ with respect to $V_{j+1}: V_{j+1}=V_j\bigoplus W_j$,
then
\begin{eqnarray}
L^2({\bf R})=\overline{V_0\displaystyle\bigoplus^\infty_{j=0} W_j},
\end{eqnarray}

As a result the solution of equations (1)-(5) has the 
following mul\-ti\-sca\-le\-/mul\-ti\-re\-so\-lu\-ti\-on decomposition via 
nonlinear high\--lo\-ca\-li\-zed eigenmodes 
{\setlength\arraycolsep{1pt}
\begin{eqnarray}\label{eq:z}
\Psi(s,x)&=&\sum_{(i,j)\in Z^2}a_{ij}U^i(x)V^j(s),\\
V^j(s)&=&V_N^{j,slow}(s)+\sum_{l\geq N}V^j_l(\omega^1_ls), \quad \omega^1_l\sim 2^l \nonumber\\
U^i(x)&=&U_M^{i,slow}(x)+\sum_{m\geq M}U^i_m(\omega^2_mx), \quad \omega^2_m\sim 2^m, \nonumber
\end{eqnarray}}
which corresponds to the full multiresolution expansion in all underlying time/space 
scales.
Formula (\ref{eq:z}) gives us expansion into the slow part $\Psi_{N,M}^{slow}$
and fast oscillating parts for arbitrary N, M.  So, we may move
from coarse scales of resolution to the 
finest one for obtaining more detailed information about our dynamical process.
The first terms in the RHS of formulae (11) correspond on the global level
of function space decomposition to  resolution space and the second ones
to detail space. In this way we give contribution to our full solution
from each scale of resolution or each time/space scale or from each nonlinear eigenmode.
This functional space decomposition corresponds to exact nonlinear
eigenmode decompositions.
It should be noted that such representations 
give the best possible localization
properties in the corresponding (phase)space/time coordinates. 
In contrast with different approaches formulae (11) do not use perturbation
technique or linearization procedures 
and represent dynamics via generalized nonlinear localized eigenmodes expansion.  
So, by using wavelet bases with their good (phase) space/time      
localization properties we can construct high-localized coherent structures in      
spa\-ti\-al\-ly\--ex\-te\-nd\-ed stochastic systems with collective behaviour.

\section{MODELLING}

Multiresolution/multiscale representations for the solutions of equations from part 2
in the high-localized bases/eigenmodes
are demonstrated on Fig.~1--Fig.~3.
This modelling demonstrates the appearance of stable patterns formation from
high-localized coherent structures.
On Fig.~1 we present contribution to the full expansion from level 1 
of decomposition (11). Fig. 2, 3 show the representations for full solutions, constructed
from the first 6 eigenmodes (6 levels in formula (11)). Figures 2, 3 show  stable pattern formation
based on high-localized coherent structures.  

\begin{figure}[htb]
\centering
\includegraphics*[width=65mm]{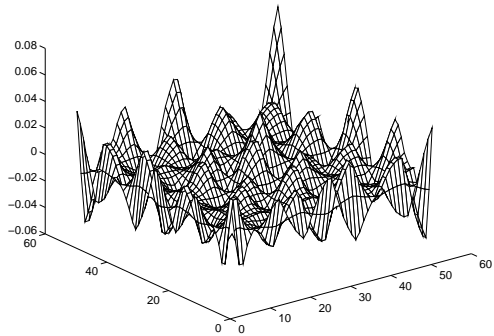}
\caption{Eigenmode of level 1.}
\end{figure}
\begin{figure}[htb]
\centering
\includegraphics*[width=65mm]{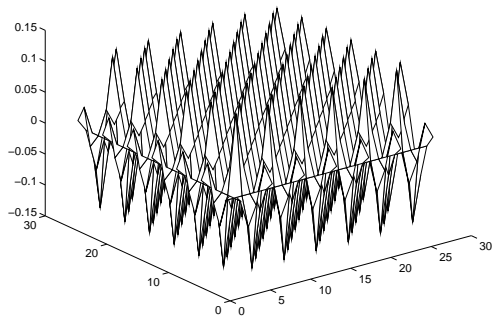}
\caption{Appearance of coherent structure.}
\end{figure}
\begin{figure}[htb]
\centering
\includegraphics*[width=65mm]{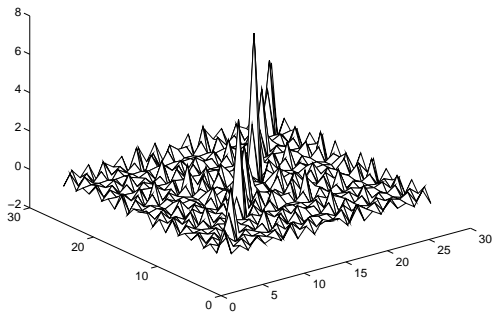}
\caption{Stable pattern 1.}
\end{figure}
\newpage

\section{ACKNOWLEDGMENTS}

We would like to thank The U.S. Civilian Research \&  Development Foundation (CRDF) for
support (Grants TGP-454, 455), which gave us the possibility to present our nine papers during
PAC2001 Conference in Chicago and Ms.Camille de Walder from CRDF for her help and encouragement.

\end{document}